\documentclass[aps,pra,superscriptaddress,nofootinbib,notitlepage,showpacs,floatfix,twocolumn]{revtex4-1}

\usepackage[utf8]{inputenc}
\usepackage[english]{babel}
\usepackage{graphicx}
\usepackage{mathptmx}

\newcommand\equalhat{\let\savearraystretch\arraystretch\renewcommand\arraystretch{0.3}
					\begin{array}{c}\stretchto{ \scalerel*[\widthof{=}]{\wedge}{\rule{1ex}{3ex}}}{0.5ex}\\ 
					= \end{array}\let\arraystretch\savearraystretch
					}

\begin{document}
\title{Designing Adiabatic Quantum Optimization: \\ A  Case Study for the Traveling Salesman Problem}
\author{Bettina Heim}
\affiliation{Theoretische Physik, ETH Zurich, 8093 Zurich, Switzerland}
\affiliation{Quantum Architectures and Computation Group, Microsoft Research, Redmond WA}
\author{Ethan W. Brown}
\affiliation{Theoretische Physik, ETH Zurich, 8093 Zurich, Switzerland}
\affiliation{Mindi Technologies Ltd. 71-74 Shelton Street, Covent Garden, London, UK, WC2H 9JQ}\author{Dave Wecker}
\affiliation{Quantum Architectures and Computation Group, Microsoft Research, Redmond WA}
\author{Matthias Troyer}
\affiliation{Theoretische Physik, ETH Zurich, 8093 Zurich, Switzerland}
\affiliation{Quantum Architectures and Computation Group, Microsoft Research, Redmond WA}
\begin{abstract}
    With progress in quantum technology more sophisticated quantum annealing devices are becoming available. While they offer new possibilities for solving optimization problems, their true potential is still an open question.
    As the optimal design of adiabatic algorithms plays an important role in their assessment, we illustrate the aspects and challenges to consider when implementing optimization problems on quantum annealing hardware based on the example of the traveling salesman problem (TSP).
    We demonstrate that tunneling  between local minima can be exponentially suppressed if the quantum dynamics are not carefully tailored to the problem.
%     A mapping of a TSP to a spin glass system that allows for a fast and extensive exploration of the solution space on the other hand introduces unwanted low energy states. Eliminating these states requires introducing additional constraints that in practice cannot successfully be implemented on analog devices. 
    Furthermore we show that inequality constraints, in particular, present a major hurdle for the implementation on analog quantum annealers. We finally argue that programmable digital quantum annealers can overcome many of these obstacles and can -- once large enough quantum computers exist -- provide an interesting route to using quantum annealing on a large class of problems. 
\end{abstract}
\maketitle

%%%%%%%%%%%%%%%%%%%%%%%%%%%%%%%%%%%%%%%%%%%%%%%%%%%%%%%%%%%%%%%%%%%%%%%%%

While quantum computing has been a long-standing topic of interest among scientists, it  has recently become the focus of public discussions as well. Its potential to be more powerful than any classical device for some applications and in particular claims that it can revolutionize the way hard optimization problems are solved \cite{vanDamIEEE} has also piqued the interest of industry. Quantum technology is maturing to the point where, for specially selected problems, it can compete with classical computers. 
Particularly, quantum annealing devices -- performing quantum optimizations by slowly evolving toward a target Hamiltonian -- and their potential have been a recent source of controversy. 
%A quantum annealer, solves optimization problems by evolving a known initial configuration at non-zero temperature towards the ground state of a Hamiltonian encoding a given problem.
For a fair assessment of the their potential it is necessary to take a close look at the real world problems they strive to solve, and how they can be implemented on a given device. Moreover, how to design such algorithms is becoming increasingly relevant as more and more sophisticated models are starting to become available \cite{ExperimentalRealization}.
In this paper we address factors that determine the performance of quantum annealing algorithms and formulate guidelines for their development. 

Quantum annealing \cite{idea_of_qa,Finnila1994,Kadowaki1998,qareview} strives to find the ground state of a target Hamiltonian $H_P$ by starting in the ground state of an easy to solve driver Hamiltonian $H_D$ and then gradually evolving the system towards the more complex target one.
	The gradual change of the Hamiltonian is described by two monotonic functions $A$ and $B$, with $A(0) = 1$, $B(0) = 0$ and $A(T) = 0$, $B(T) = 1$, such that 
	the Hamiltonian at a time $t$ is given by 
	\begin{equation}
	H(t) = A(t)H_{D} + B(t)H_{P} \ \ \textrm{ for } t \in [0,T].
	\end{equation}
	A common choice is $A(t) = 1 - t/T$ and $B(t) = t/T$. 
Ideally, the system remains in the ground state of the instantaneous Hamiltonian during the evolution. The quantum adiabatic theorem gives a sufficient condition for this to be the case.
In analog devices, thermal as well as quantum fluctuations can excite the system, making quantum annealing an approximate solver that will generally find states close to but not necessarily the exact ground state.
	
	In order to solve an optimization problem by annealing, its solution needs to be encoded into the ground state of the target Hamiltonian. With quantum annealing being an approximate solver, it is preferable that in fact all low energy states correspond to solutions that are close to optimal - and to only those.
	Since the commutation relation between $H_{D}$ and $H_{P}$ determines the dynamics during evolution, the chosen encoding additionally has to permit the use of a simple enough to implement driver that allows for fast transitions between potential solutions. 

While in principle it is possible to solve an arbitrary problem on an annealing device, its quantum nature as well as architectural limitations 
	impose restrictions on the cost functions and possibly constraining conditions that can be realized.
	Optimally implementing a given problem thus requires a well chosen mapping onto a suitable target Hamiltonian. The choice of this mapping significantly influences the performance of the algorithm and its scaling with problem size. 
Whether or not a problem can be solved efficiently by annealing thus depends on both the available hardware and the chosen algorithm. 
We discuss the issues that need to be considered when designing specialized quantum hardware 
and illuminate the challenges and pitfalls of adiabatic quantum computing by examining the case of the traveling salesman problem (TSP). We then show that many of these problems can be overcome on gate-model quantum computers.
%Even though a faithful encoding of a TSP into physical hardware is possible, we show that such an encoding is not well suited to the mechanisms exploited by quantum annealing and bound to give a poor performance. We test a mapping whose performance is less likely to suffer with increasing problem size and is compatible with dynamics that allow for a fast an extensive exploration of the solution space. 
%In contrast to state of the are classical algorithm based on such an encoding, we show that it is unlikely to succeed on analog quantum devices as it requires an iterative procedure implementing an increasing number of inequality or competing constraints. 

%%%%%%%%%%%%%%%%%%%%%%%%%%%%%%%%%%%%%%%%%%%%%%%%%%%%%%%%%%%%%%%%%%%%%%%%%

{\it Mapping the TSP to an annealing problem} --
Given $N$ cities % $1,\dots , N$ 
and distances $d_{ij}$ between them, the task of the traveling salesman problem is to find the shortest possible roundtrip that visits each city exactly once. %This problem has to be mapped to that of finding the ground state of a Hamiltonian that can be realized on a quantum annealer. Current devices provide local fields and tunable two-site couplings between adjacent qubits, implying that the target Hamiltonians have to take an Ising form. 
Since current devices provide only local fields and tunable two-site couplings between adjacent qubits, any target Hamiltonian has to correspond to an Ising spin glass.
The first step is to represent every possible valid roundtrip as a spin configuration. The straightforward encoding is to associate each roundtrip with a permutation matrix $a_{ik}$, where $a_{ik}=1$ if the $i$-th city is visited at time $k$ of the tour, and zero otherwise \cite{Lucas14}.  With the mapping $a_{ik} = (1 - \sigma_{ik}^z)/2$ the Hamiltonian can be formulated in terms of quantum spin variables. 
	%Choosing the starting point of the roundtrip, $(N-1)^2$ qubits are sufficient to represent these matrix elements.
	%A total of $(N-1)^2$ qubits are sufficient to represent these matrix elements, since the starting point of the roundtrip can be chosen arbitrarily.
We then need to ensure that the ground state corresponds to the encoding of the shortest roundtrip. 
Minimizing the tour length given by the Hamiltonian
	\begin{equation}
		H_l = \sum_{i, j, k} d_{ij} a_{ik}a_{jk+1} %\\
	\end{equation}
subject to the constraints $\sum_i a_{ik}  = 1 \ \forall k$ and $\sum_k a_{ik} = 1 \ \forall i$
accomplishes our goal.
%where we defined the binary observable $a_{ik} = (1 + \sigma_{ik}^z)/2$. % acting on spin $s_{ik}$.
These two requirements guarantee that $(M_{ij})$ is indeed a permutation matrix. They can be implemented by constraint terms
	\begin{equation}
		\label{std_constraints}
		H_c = \sum_{i} \left[\left (1 - \sum_{j} a_{ij} \right)^2 + \left(1 - \sum_{j} a_{ji}\right)^2 \right],
	\end{equation}
	which add an energy penalty to states violating them. The ground state of the Hamiltonian $H_{P} = H_l + \eta H_c$, for $\eta \geq \max \{ d_{ij}/2 \}$, therefore provides the desired TSP solution. 
Note that for an $N$-city TSP we in principle require $(N-1)^2$ qubits and $(N-2)(N-1)^2 + N^2(N-1)$ couplers. 
% we have (N-1)*2 spins, each with 2(N-2) neighbors, giving (N-2)(N-1)^2 couplings needed for the constrains,
% plus the couplings needed for implementing the cost function, which are 2*N times the N(N-1)/2 pairs of cities
Given a typical QA architecture with a small bounded number of couplers per qubit, one will rather need ${\mathcal O}(N^3)$ qubits.

\begin{figure}
	\includegraphics[width=\columnwidth]{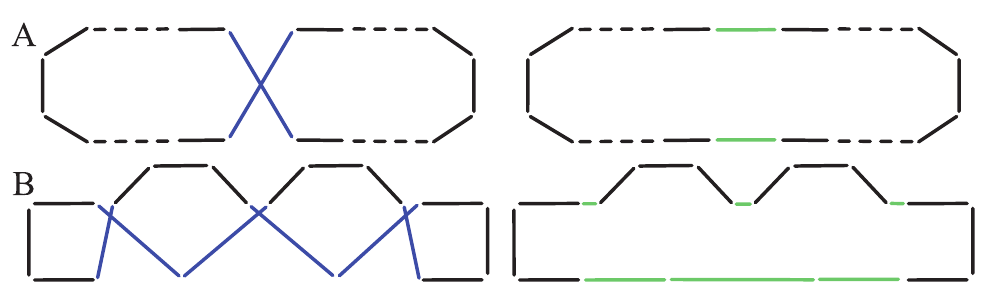}
	\caption{A) A crossing requiring up to $4\lfloor N/4\rfloor$ single spin flips to resolve for a permutation mapping, and only $4$ for a symmetric TSP represented by a graph mapping. B) Worst case for $N = 18$, $r = 3$: Using a permutation mapping, resolving $r$ crossing requires up to $2\big(N - \lceil (N-(r-1))/(r+1)\rceil\big)$ single spin flips.}
	\label{tours}
\end{figure}

\vspace{0.3cm}
{\it The quantum driver Hamiltonian} --
The next step is to find a quantum driver Hamiltonian $H_{D}$ determining the dynamics of the annealing process, that provides an efficient near-adiabatic evolution towards $H_{P}$ without ending up in an excited state. The usual choice is a transverse field term $H_x = \Gamma \sum_{i,j} \sigma_{ij}^x$, which induces single spin flips. Before contemplating more complex alternatives it is useful to understand the influence of $H_{D}$ on the annealing efficiency. Consider the probability to transition between two tours of similar length, as shown in Fig.~\ref{tours}A. This transition can be performed by a so-called 2-opt update \cite{2opt_moves, martonak2004TSPannealing}, which is a common and very efficient primitive move in classical heuristics. Using the above mapping this, however, requires to update $m={\mathcal O}(N)$ variables, since we have to change the order in which half of the cities are visited. 
The probability to transition between these two tours towards the end of the annealing process is thus -- in leading order -- proportional to $\big(\Gamma/\Delta)^{m}$ \cite{SM}, where $\Delta$ is the scale associated with the barriers between the two solutions. % otherwise I would have to recalculate in detail, may depend on the derivative
A simple crossing, as shown in Fig. \ref{tours}A, is therefore difficult to resolve since the transition probability is exponentially suppressed (in the problem size $N$) compared to classical heuristics that can directly implement a 2-opt update.
%Resolving $r$ crossings by single spin flips involves close to $N$ moves in the worst case (see Fig.\ref{tours}B).

We thus see that the choices of mapping $H_{P}$ and $H_{D}$  affect which updates to a configuration are efficiently realized during quantum annealing, and this directly and significantly impacts performance. The above exponential slowdown might be avoided by a better choice of $H_{D}$ or $H_{P}$. Following the first route we could opt to permute several cities using multi-qubit couplers. 
While resolving a crossing may still entail $\mathcal{O}(N)$ steps and the exponential suppression remains, this may nevertheless significantly improve transition probabilities by avoiding high energy intermediate configurations that violate constraints. In fact, such kinetics could allow sampling of only viable TSP solutions, which would render the constraints of Eq. (\ref{std_constraints}) unnecessary, and thereby simplify the energy landscape that needs to be explored \cite{DriverHamiltonian}. However, the pairwise exchange of all two-city pairs requires $\mathcal{O}(N^4)$ four-spin couplers, which is infeasible for all but the smallest problems.
 
{\it An improved mapping -- }
In order to design a mapping that allows for an efficient realization of 2-opt (or more generally $k$-opt) moves in the quantum annealer we associate  $a_{ij}=a_{ji}$ with the undirected edges between cities $i$ and $j$. Using a cost function
\begin{equation}
 		H'_l = \sum_{i, j} d_{ij} a_{ij},
\end{equation}
TSP solutions are subject to the constraint that the set of edges with $a_{ij}=1$ form a valid tour. With this mapping a 2-opt update only requires the flipping of $m=4$ spins. More general $k$-opt move requires just $m=2k$ flips, independent of the problem size. Such a mapping thus avoids the exponential slowdown of the previous one.

While the the number of required qubits seems to be comparable at $N(N-1)/2$, this number can be substantially reduced by truncating the set of considered edges. Along the optimal tour, cities are connected almost exclusively to nearby cities. In fact, the probability of connecting to the $l$-th farthest city decreases exponentially with $l$ for random problems instances. We can thus truncate the set of considered edges originating at a city to a small number of $L$ closest cities. This substantially reduces the number of required qubits to $NL/2 = {\mathcal O}(N)$.

{\it Implementing the constraints --} Closed tours can be enforced by adding a constraint term
\begin{equation}
H_c' = \sum_i \left(2 - \sum_{j \neq i} a_{ij}\right)^2,
\label{eq:const}
\end{equation}
which enforces that each city is connected to two edges. These constraint terms require $\mathcal{O}(NL^2)$ 2-qubit couplers, substantially less than the $\mathcal{O}(N^3)$ terms required for the first mapping.
While this term enforces a configuration consisting of closed loops where each city is visited exactly once, it does not in fact enforce that all visited cities belong to the same loop: the tour can break up into disjoint {\em subtours}. Depending on the specific variant of the TSP this may or may not be desired -- one may, for example, want to know if using multiple salesmen is preferred. However, for randomly generated problems many of the subtours are not particularly interesting. Evaluating 100 random problems with $N=12$ and uniformly distributed cities in a 2D-plane using CPLEX \cite{cplex,Optima98} shows that the ground state of 75\% of the instances splits into subtours and a majority of these subtours contain only three cities. For larger problem sizes, it is likely that here too, we will frequently obtain solutions consisting of a large number of subtours containing only a small number of cities. 
 
Directly enforcing a single closed tour would require $N$-qubit coupling terms and is unrealistic. The standard procedure to avoid such undesired states is to iteratively  add terms that penalize the {\em specific} subtour breakups encountered during the optimization. Given a breakup into, e.g. two sets of cities $\mathcal A$ and $\mathcal B$, we add an inequality constraint of the form
	\begin{equation}
		\label{subtour_constraint_inequality}
		\sum_{i \in \mathcal{A}} \sum_{j \in \mathcal{B}} a_{ij} > 0.
	\end{equation}
	
	Unfortunately, such an inequality constraint is hard to implement with two-qubit couplings in an Ising model quantum annealer. Approximating the step function of an inequality by a $k$-th order polynomial requires implementing $\mathcal{O}(N^{2k})$ $k$-spin couplings. 
		
Luckily, an evaluation using CPLEX \cite{cplex} shows that for the ground states of our instances there are very few required connections; around 94\% of disconnected subtours should have merely two connections with each other and the remaining 6\% should form four connections. % though these are distributed across 15\% of the systems.
For these a simple quadratic energy penalty
	\begin{equation}
		\eta' \left(C-\sum_{\in \mathcal{A}} \sum_{j \in \mathcal{B}} a_{ij}\right)^2
		\label{subtour_constraints}
	\end{equation}
	with a constant $C=2$ to favor two connections or $C=3$ to equally favor two and four connections would be sufficient. Such constraint terms increases the number of couplings to  $\mathcal{O}(N^2L^2)$, which is still a better scaling than in the original mapping.	
The algorithm to obtain an estimate for the TSP solution then consists of first annealing the minimally constrained system described by $H_{P} = H_l' + \eta H_c'$. If the best solution found splits into subtours, we add additional  constraints (\ref{subtour_constraints}) before repeating the annealing. This procedure is repeated until a solution consisting of a single closed tour is found.

{\it Simulation results --} 
We analyzed the effectiveness of this algorithm by numerical simulations on problems with $N=8$, 12 and 16 cities. We focus our discussion here on the main results for the case $N=12$. We investigated 100 random TSPs with the cities uniformly distributed on a square.

We start by testing the subtour suppression strategy using the MIQP solver of CPLEX \cite{cplex}. To avoid any complications due to competing constraints, we first analyze the performance of the outlined algorithm when choosing $C = 2$ for all iterations. This should enforce the correct behavior for the majority of instances where only two connections between subtours are required. Indeed, after one iteration almost all subtours require merely two connections with only one needing four, and after just two iterations the optimal TSP solution is found for 95\% of these systems.

\begin{figure}
	\includegraphics[width=\columnwidth]{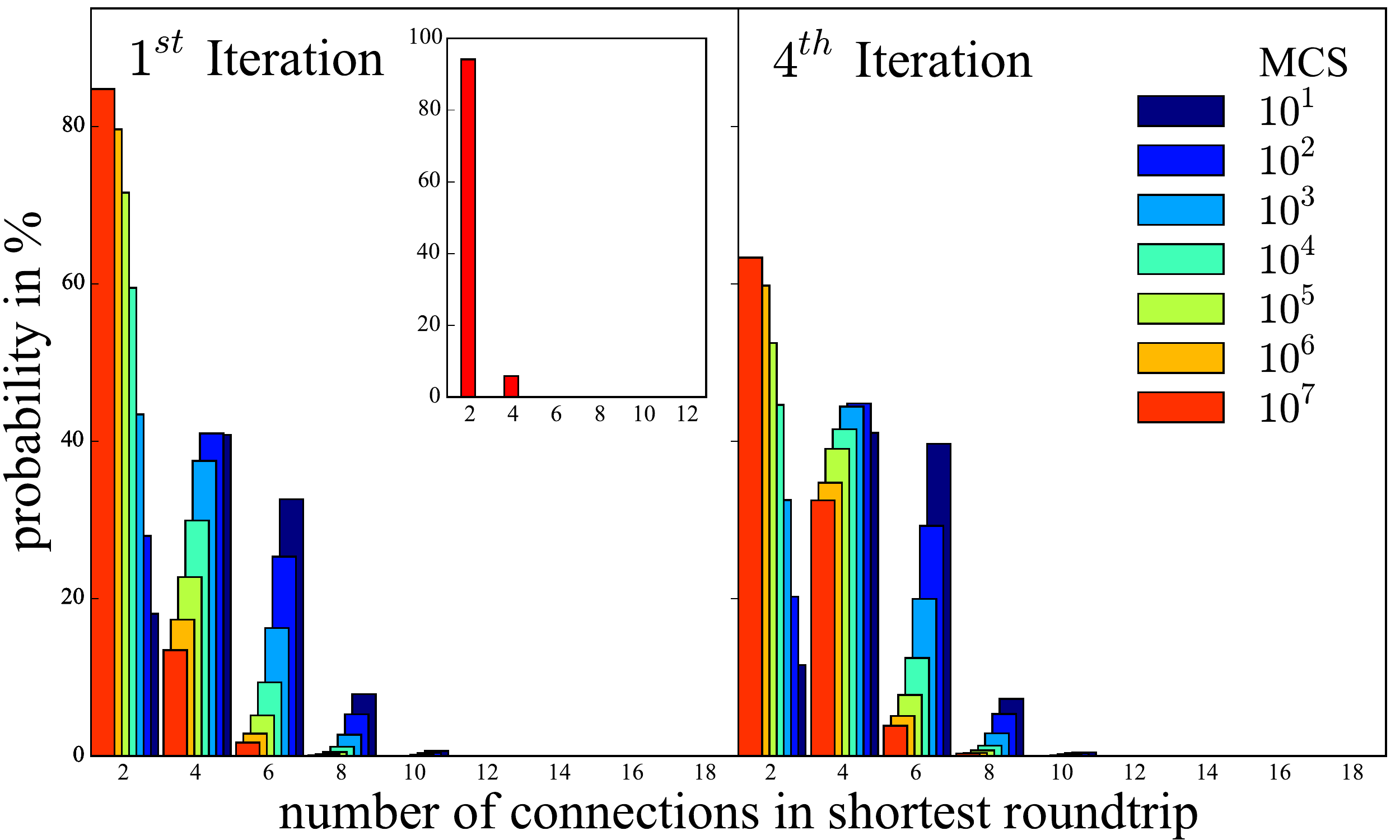}	
	\caption{Distribution of the number of connections that a subtour found by annealing should have during the first and forth iteration in order to be consistent with the TSP solution. As the state after annealing is generally an excited state, the number of connections can be quite high even for problems where the ground state subtours require only very few connections - even more so the farther we are from the ground state. The legend denotes the number of Monte Carlo steps (MCS) used for annealing in both panels. The inlay in the ``$1^{st}$ Iteration'' panel shows the distribution for the ground state subtours obtained by an exact solver.}
	\label{connection_statistics}
\end{figure}

Even though iteratively adding constraints works reasonably well with exact solvers, we found that it fails with heuristic solvers, such as QA, simulated QA (SQA), or classical simulated annealing (SA). We show SA results but expect the observations to carry over to SQA and QA. The algorithm succeeds in finding the TSP solution only in very few cases. 
The reason for this failure is the limited probability of finding the absolute minimum. Unfortunately, this does not simply translate into a larger number of repetitions before the algorithm terminates. 
Contrary to the ground states, a significant number of the subtours found by annealing should have more than two connections in the TSP solution. 
As can be seen in Fig. \ref{connection_statistics}, a poor annealing performance significantly reduces the chance of introducing an appropriate set of constraints. 
Since enforcing the \emph{wrong} number of connections - that is one inconsistent with the TSP solution - during any one repetition implies that the roundtrip obtained at the end of our algorithm is not of minimal length, the success probability of our algorithm decreases exponentially with the number of iterations. 

In an effort to mitigate the detrimental effects resulting from the uncertainty about the required number of connections %found subtours should form % of a limited annealing performance 
one could pursue several strategies. Adding a penalty function that has multiple minima, e.g. at $C=2$ and $C=4$ requires   $\mathcal{O}(N^8)$ four-spin couplings and is thus not likely to be implementable in the near future.  Instead, one might try to choose $C = 3$ in order to equally favors two or four connections, given that an even number of connections is enforced. As the obtained subtours can contain a similar set of cities for several iterations, the ratio $\eta/\eta'$ then needs to be successively increased with each iteration; otherwise the ground state configuration corresponds to broken tours with three connections between subsets of cities. This creates an unfavorable and very rough energy landscape, where an annealer has barely any chance of finding the ground state. 

A potential alternative is to use slack variables ${s_1\dots s_m}$, ${s_k \in \{0,1\}}$, for each subset $\mathcal A$ of $m$ cities forming a subtour. One can then implement soft constraints by introducing energy penalties 
\begin{equation}
\eta' \bigg( \sum\limits_{i\in\mathcal{A}} \sum\limits_{j \not\in\mathcal{A}} a_{ij} - \sum\limits_{k=1}^{m} 2ks_k \bigg) ^2 + \eta'' \bigg(\sum\limits_{k=1}^{m} s_k - 1\bigg)^2.%	
\end{equation}
Engineering a suitable energy landscape, however, poses similar challenges, and transitions between solutions with a different number of connections can be heavily suppressed. 

We thus conclude that analog quantum annealing devices are unlikely to be of interest as TSP solvers in the near future.

\

\begin{figure}
	\includegraphics[scale=0.35]{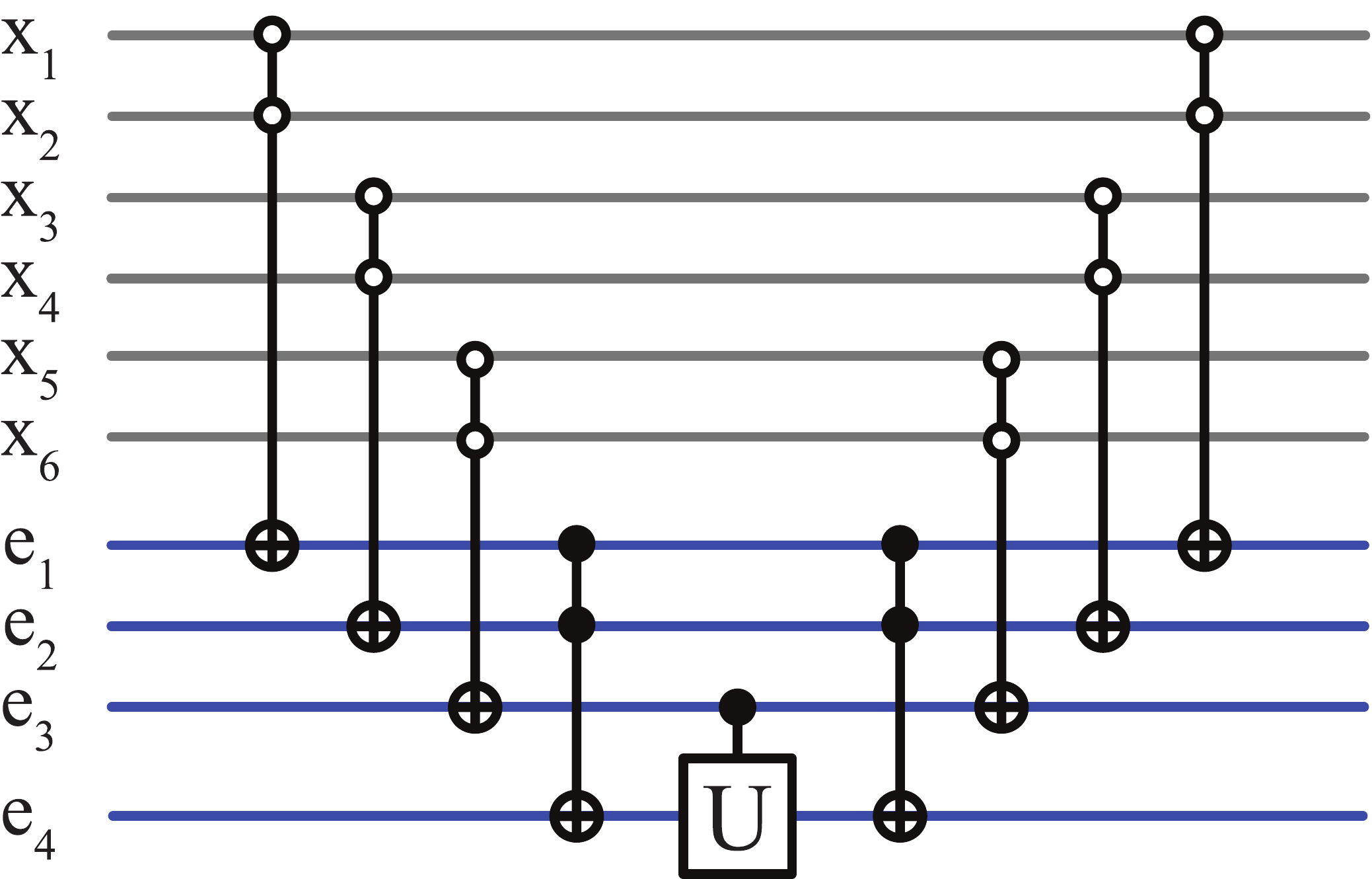}	
	\caption{Circuit implementing an energy penalty if a certain subset of cities is disjoint from the rest (inequality constraint (\ref{subtour_constraint_inequality})). The qubits $x_1\dots x_m$ represent all possible connections between the subset and the other cities, the qubits $e_1 .. e_{m-2}$ are additional ancilla qubits initialized to $|0\rangle$ (the graphic shows $m=4$). The $2(m-2)$ Toffoli gates can be executed in $\mathcal{O}(\log m)$ time. Open circles denote conditioning on the connections $x_i$ not being part of the current tour configuration. The unitary $U$ is a phase gate that implements the propagator corresponding to an energy penalty $\eta'$ during one step of the annealing process by adding a phase $\exp (-i B(t)\eta'\Delta_t/\hbar)$ if the qubit is set.}
	\label{digital}
	%\includegraphics[width=\columnwidth]{graphics/circuit2.eps}	
	%\caption{Recursion step of the circuit used to eliminate all subtours simultaneously. The circuit results in the ancilla qubit $a_1$ being set if and only if there is a closed subtour of length $k+2$. The red gates are necessary only to minimize the number of qubits. While this circuit illustrates the principle, using more ancillas allows one to reduce the scaling of the number of gates with system size and is thus more reasonable. }
	%\label{circuits}
\end{figure}

%\begin{comment}
%Even if we have a fairly accurate idea of how many connections a found subtour should form, eliminating a large number of them by adding energy penalties proves difficult. The reason behind this being that introducing generic constraints is extremely difficult without negatively affecting the energy landscape. Strong constraints, that is a large scaling factors $\eta$ and/or $\eta'$, cause energy barriers that are hard to bypass. Furthermore, there is a good change that they lay close to the ground state, inducing the need to pass them to find the optimal solution. Using two-opt moves would allow to bypass those barriers much easier as only part of the configuration space is sampled and the width of these barriers is therefore effectively decreased, allowing for higher tunneling probabilities. However, implementing arbitrary two-opt moves is difficult and costly. Weak constraints, on the other hand, flatten the energy landscape, thereby making it harder to find the optimal solution. Thinking that such a landscape at least allows find a configuration close to the optimal solution would be wrong, too, as energetically close states of the heavily engineered Hamiltonian are not necessarily close with respect to the original spectrum. 
%
%Whereas we are forced to eliminate subtours one by one on an analog device ... 
%\end{comment}

%{\color{red} fixme: this conclusion needs to be updated to be consistent with the previous section} \\
{\it Digital Quantum Annealing --}
Virtually all of the above mentioned issues can be remedied by a ``digital'' implementation on a gate-model quantum computer that simulates the time evolution of quantum annealing by splitting the propagation into discrete time steps $\Delta_t$ \cite{Barends2015, Lloyd96}. This has several advantages: Quantum error correction removes calibration errors. The flexibility offered by a programmable universal quantum computer offers more choices of quantum dynamics, including 2-opt moves. Embedding the program into a specific hardware graph imposes at most linear overhead in runtimes, opposed to potentially exponential slowdown of quantum tunneling due embedding into a system with low connectivity in an analog approach. The inequality constraint (\ref{subtour_constraint_inequality}) can now be implemented without heavy approximations. Finally, all penalty terms in the cost function can be implemented much more efficiently, reducing the scaling of the number of qubits with problem size.

Implementing a constraint $\sum_i^m x_i=a$ as a quadratic function  $(\sum_i^m x_i-a)^2$ requires $m^2/2$ couplers, which results in $\mathcal{O}(m^2)$ qubits assuming limited connectivity. The same constraint can be implemented in a digital simulation as a phase rotation conditioned on whether the constraint is satisfied or not. %, i.e. a phase gate applied to a qubit containing the result of $\delta_{\sum_i^Nx_i,a}$, as shown in Fig. \ref{digital} . 
Using just $\mathcal{O}(m)$ qubits this can be implemented in time   $\mathcal{O}(\log m)$ (see Fig. \ref{digital}). With this approach the constraint (\ref{eq:const}) requires only $\mathcal{O}(N^2)$ instead of $\mathcal{O}(N^3)$ qubits and the cost for the constraint   (\ref{subtour_constraint_inequality}) is $\mathcal{O}(N^2)$. 
Furthermore, a more even energy landscape allows for better annealing performance.

Simulating quantum annealing using QMC simulations on a classical computer profits from the same advantages of digital quantum annealing and may thus be a promising route to explore.

{\it Conclusion --} The traveling salesman problem demonstrates many important aspects to consider in the design of both adiabatic quantum algorithms and specialized hardware.  A so far under appreciated aspect is that quantum dynamics has to be an important consideration  in designing the mapping of an application problem to Ising spin variables. We argued that using transverse fields (or any other local term) for the quantum dynamics incurs an exponential slowdown in the standard faithful mapping of TSP to Ising spins, compared to efficient 2-opt updates. We thus considered an alternative mapping, which avoids this slowdown. 
The improved dynamics for this alternative mapping comes at the cost of requiring additional constraints to prevent a breakup into subtours. We found that the limitation to quadratic penalty functions in an Ising model constitutes a problem. In particular the need for inequality constraints presents a major hurdle for the implementation on an analog quantum annealing device. 

These problems can be solved by considering a digital implementation of quantum annealing on a universal quantum computer simulating QA, or on a classical computer implementing a QMC version of QA. The programmability of the digital computer allows efficient implementation of a large class of cost functions and penalty terms. Furthermore, the scaling of the required number of qubits is quadratically improved from  $\mathcal{O}(N^4)$ to   $\mathcal{O}(N^2)$ (or from $\mathcal{O}(N^2L^2)$ to $\mathcal{O}(NL)$  when using a cutoff $L$ for  the number  of neighboring cities considered). 

We thus see digital quantum annealers as a  promising route to quantum optimization, also because they allow more tailored types of quantum dynamics to be programmed and -- with error correction -- solve the calibration and error problems of analog devices.

We thank Donjan Rodic and Ilia Zintchenko for helpful discussions. MT acknowledges hospitality of the Aspen Center for Physics, supported by NSF grant PHY-1066293. This work has been supported by the Swiss National Science Foundation through the NCCR QSIT and by ERC Advanced Grant SIMCOFE.
This paper is based upon work supported in part by ODNI, IARPA via MIT Lincoln Laboratory Air Force Contract No. FA8721-05-C-0002. The views and conclusions contained herein are those of the authors and should not be interpreted as necessarily representing the official policies or endorsements, either expressed or implied, of ODNI, IARPA, or the U.S. Government. The U.S. Government is authorized to reproduce and distribute reprints for Governmental purpose not-withstanding any copyright annotation thereon.

\bibliography{biblio.bib}{}
\end{document}